\documentclass[aps,prl,twocolumn,showpacs,amsmath,floatfix]{revtex4-1}

\usepackage{graphicx}
\usepackage{braket}

\newcommand{\MoS}[1]{MoS\ensuremath{{}_{#1}}}
\newcommand{\GWo}{}
\def\GWo/{GW$_0$}
\newcommand{\GoWo}{}
\def\GoWo/{G$_0$W$_0$}
\newcommand{\scGWo}{}
\def\scGWo/{sc-GW$_0$}

\begin{document}


\title{Optical Spectrum of MoS${}_2$:  Many-body Effects and Diversity of Exciton States}

\author{Diana Y. Qiu}
\affiliation{Department of Physics, University of California at
Berkeley, California 94720}
\affiliation{Materials Sciences Division, Lawrence Berkeley
National Laboratory, Berkeley, California 94720}

\author{Felipe H. da Jornada}
\affiliation{Department of Physics, University of California at
Berkeley, California 94720}
\affiliation{Materials Sciences Division, Lawrence Berkeley
National Laboratory, Berkeley, California 94720}

\author{Steven G. Louie}
\email[Email: ]{sglouie@berkeley.edu}
\affiliation{Department of Physics, University of California at
Berkeley, California 94720}
\affiliation{Materials Sciences Division, Lawrence Berkeley
National Laboratory, Berkeley, California 94720}

\date{\today}

\begin{abstract}
We present first-principles calculations of the optical response of monolayer molybdenum disulfide employing the GW-Bethe Salpeter equation (GW-BSE) approach including self-energy, excitonic, and electron-phonon effects. We show that monolayer MoS2 possesses a large and diverse number of strongly bound excitonic states with novel k-space characteristics that were not previously seen experimentally or theoretically. The absorption spectrum is shown to be dominated by excitonic states with binding energy close to $1$~eV and by strong electron-phonon broadening in the visible to ultra-violet range. Our results explain recent experimental measurements and resolve inconsistencies between previous GW-BSE calculations.
\end{abstract} 

\pacs{73.22.-f, 71.35.-y, 78.67.-n}

\maketitle


Atomically thin structures of layered materials, such as graphene and transition-metal dichalcogenides, are of great interest in condensed matter physics due to their remarkable electronic properties and applicability in nanoscale devices\cite{mak10,splendiani10,radis11,yoon11}. In this category of two-dimensional (2D) systems, monolayer molybdenum disulfide (\MoS2) has recently gained attention for combining an electron mobility comparable to graphene with a finite energy gap\cite{radis11}. Unlike its bulk form, which is an indirect gap semiconductor, monolayer \MoS2 has a direct gap at the $K$ and $K'$ points in the Brillouin zone, as a result of inversion symmetry breaking in its honeycomb lattice structure\cite{mak10,splendiani10}.

As a 2D material, \MoS2 is expected to have strong excitonic effects, which will influence its optical properties. Thus, it is important to accurately calculate and understand the many-electron properties of \MoS2. One common approach to compute quasiparticle band structure and optical response including electron-electron interactions and excitonic contributions is the GW plus Bethe Salpeter equation (GW-BSE) approach\cite{hybertsen86,rohlfing98,benedict98,reining98}. This method has been applied with success to a wide variety of materials\cite{louiechapter06}, including systems with reduced dimensionality such as graphene\cite{yang09} and nanotubes\cite{spataru04,deslippe07}. Recent GW calculations of the quasiparticle band structure of \MoS2, however, have reported conflicting findings. Previous work based on the quasiparticle self-consistent GW (QSGW)\cite{lambrecht12} and one-shot \GoWo/\cite{ashwin12} methods conclude that \MoS2 has a direct band gap, as also deduced from optical absorption and photoluminescence experiments\cite{mak10,splendiani10}. On the other hand, a more recent work using the self-consistent (sc) \GWo/ approach claims that \MoS2 has an indirect gap at the \GoWo/ level, while self-consistently updating the Green's function $G$ changes it to a direct gap\cite{shi13}. There is also significant disagreement on the optical excitations calculated at the GW-BSE level, with the BSE results based on the previous \GoWo/ calculation\cite{ashwin12} reporting only two  bound exciton peaks while the results based on the \scGWo/ work find one extra peak at $\sim 2.5$~eV, with the positions of all the peaks being quite sensitive to the sampling of the Brillouin zone\cite{shi13}.

Since the same GW-BSE formalism was used, these disagreements are not a shortcoming of the GW-BSE method itself, but rather an indication of a computationally challenging system. For instance, it was shown that a large number of bands ($N_b$) and a high energy cutoff ($E_S$) for the dielectric matrix are required whenever the quasiparticle wave functions involve localized states or when there is a significant difference between the characters of the valence band maximum (VBM) and the conduction band minimum (CBM) states\cite{shih10}. These features are precisely present in \MoS2, since the VBM at $K$ and $K'$ have localized Mo $d_{x^{2}-y^{2}}$ and S $p_{x}$ and $p_{y}$ characters, while the CBM has a Mo $d_{z^{2}}$ character with a small contribution from S $p_{x}$ and $p_{y}$\cite{cao12}. Another issue is that accurate description of excitonic states requires fine k-space sampling, since the excitonic states are correlated bound states in real space.

\begin{table*}
    \caption{\label{tab:gw}Comparison of quasiparticle band gap ($E_{g}^{GW}$) and effective masses at K from different studies and the respective convergence parameters used --- k-grid, dielectric energy cutoff ($E_{S}$), and number of bands ($N_{b}$)}
    \begin{tabular*}{\textwidth}{@{\extracolsep{\fill}}ccccccc@{}}
        \hline
        \hline
		 ~ & \multicolumn{1}{c}{QP Gap at K} & \multicolumn{2}{c}{Effective Masses} & \multicolumn{3}{c}{Convergence Parameters} \\
		 \cline{2-2} \cline{3-4} \cline{5-7} 
        ~  &E$_{g}^{GW}$ (eV) &m$^{*}_{e}/m_{0}$ & m$^{*}_{h}/m_{0}$ & k-grid & $E_S$ (eV) & $N_b$\\ 
        \hline
       
        Present Work (G$_{1}$W$_{0}$) & 2.84 & 0.37 & 0.21 & 12x12x1 & 476 & 6000 \\ 
        QSGW\cite{lambrecht12} & 2.76 & 0.35 & 0.44 & 8x8x2 & -- & -- \\
        \GoWo/\cite{ashwin12} & 2.82 & 0.60 & 0.54 & 6x6x1 & 272 & 96 \\
        \scGWo/\cite{shi13} & 2.80 & 0.36 & 0.39 & 12x12x1 & 300 & 197 \\
        \hline
        \hline
    \end{tabular*}
\end{table*}

In this work, we calculate the quasiparticle band structure and optical spectrum of \MoS2 from first principles, taking special care to obtain well-converged results to within 0.1~eV. We predict that the absorption spectrum of \MoS2 has very rich additional physics in the region between the energies of the first two bound exciton states and energies close to the quasiparticle gap. To explore this region, we perform a comprehensive analysis of the effect of k-space sampling on the existence and convergence of bound exciton solutions to the BSE. In particular, we discovered a series of optically active exciton states with novel characteristics (e.g., annular k-space wave functions) not seen in previous calculations, and we identify multiple excited states of the first set of bound excitons and their respective intra-exciton transition energies. These new states of diverse character are necessary for a conceptual and quantitative understanding of experimental absorption spectra\cite{mak10} and recent time-resolved photoluminescence spectra\cite{lagarde13}. Furthermore, the unusually large number and diverse character of bound excitons in \MoS2 suggest that it could be an excellent candidate for exploring inter- and intra-excitonic processes.

In this study, we first perform mean-field calculations of \MoS2 using density functional theory (DFT) in the local density approximation (LDA) using the \textsc{Quantum Espresso} code\cite{espresso}. The GW calculations are performed with the \textsc{BerkeleyGW} package\cite{jdeslip11BGW}. We calculate the dielectric matrix $\epsilon_{\mathbf{G,G'}}^{-1}(\mathbf{q})$ and the self-energy $\Sigma$ with a truncated Coulomb interaction\cite{beigi06} on a 12x12x1 k-grid to converge our quasiparticle gap to within $0.1$~eV.

\begin{figure}
    \includegraphics[width=246.0pt]{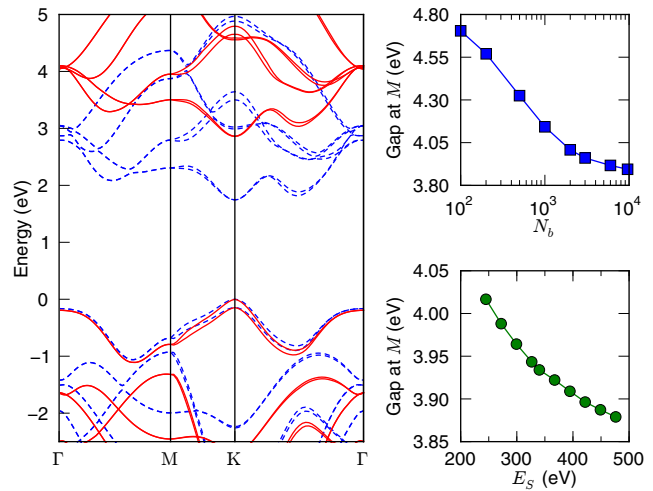}
    \caption{\label{fig:bandstructure} (Color online) Left: LDA (dashed blue curve) and GW (solid red curve) band structure of monolayer \MoS2. Top right: convergence of the GW gap at $M$ as a function of the number of empty bands included, with $E_S$ fixed at $476$~eV. Bottom right: convergence of the GW gap at $M$ as a function of the energy cutoff $E_S$ of the dielectric matrix, with $N_b$ fixed at $9{,}600$.}
\end{figure}

We find it necessary to employ a significantly higher energy cutoff of $E_S=476$~eV for the dielectric matrix $\epsilon_{\mathbf{G,G'}}^{-1}$ than those values reported in previous calculations. We also need to include $N_b=6{,}000$~bands\footnote{Throughout this paper, the number of bands will always refer to the number of spin-orbit bands, i.e., the number of bands in our spin-unpolarized calculation.}, which is more than an order of magnitude more than any previous calculation, in calculating both $\epsilon_{\mathbf{G,G'}}^{-1}$ and the self-energy $\Sigma$. With these parameters, we converge the valence and conduction bands to within better than $0.1$~eV of accuracy. We also stress that $E_S$ and $N_b$ are not two independent convergence parameters. Thus, attempting to converge $N_{b}$ with an unconverged $E_{S}$ may lead to false convergence behavior\cite{shih10}. In Fig.~\ref{fig:bandstructure}, we show the convergence behavior of the quasiparticle band gap at M, illustrating the importance of having adequate values for both $N_{b}$ and $E_{S}$. In Table~\ref{tab:gw} we compare our convergence parameters and results with those in the literature.

In Fig.~\ref{fig:bandstructure}, we compare our DFT-LDA and GW band structures. We find that \MoS2 is a direct gap semiconductor at both the LDA and \GoWo/ levels. The LDA gap is 1.74~eV and the \GoWo/ gap is 2.78~eV at both the $K$ and $K'$ points. Performing one self-consistent update for G increases the gap to 2.84~eV. An additional self-consistent step changes the gap by less than 20~meV, so we stop at at the G$_{1}$W$_0{}$ level. Spin-orbit (S-O) coupling splits the valence band at the $K$ point by 146 meV. These S-O and band gap values at K are in good agreement with a previous QSGW\cite{lambrecht12} and \GoWo/\cite{ashwin12} calculations. However, we find different values for the effective mass of the charge carriers at K. Values for quasiparticle gap and effective masses from our calculation and previous calculations are reported in Table~\ref{tab:gw}. Values for the DFT-LDA gap and S-O splitting are reported in the Supplemental Material\footnote{See Supplemental Material at [url] for additional details on the computational details, convergence parameters, inclusion of S-O coupling, and quasiparticle lifetimes.}.

\begin{table*}
    \caption{\label{tab:bse}Values from different studies: optical transition energies for peaks A, B, A', B', and C, the binding energy of peak A (E$_{b}$) and the convergence parameters --- k-point sampling and number of valence (N$_{v}$) and conduction (N$_{c}$) bands.}
    \begin{ruledtabular}
    \begin{tabular*}{\textwidth}{@{\extracolsep{\fill}}cccccccccc@{}}
    	& \multicolumn{6}{c}{Optical Transition Energies (eV)} & \multicolumn{3}{c}{Convergence Parameters} \\
		\cline{2-7} \cline{8-10}
		&A & B & A' & B' & C & E$_{b}$ & k-grid & N$_{v}$ & N$_{c}$ \\
		\hline
        Present Work (G$_{1}$W$_{0}$) & 1.88 & 2.02 & 2.20 & 2.32 &2.54 & 0.96  & 72x72x1 & 7 & 8 \\
        Absorp. Exp.\cite{mak10} & 1.88 & 2.03 & -- & -- &-- & -- & -- & -- & -- \\
        PL Exp. \cite{splendiani10} & 1.85 & 1.98 & -- & -- & -- &-- & -- & --& -- \\
        \GoWo/-BSE\cite{ashwin12} & 1.78 & 1.96 & --&-- & -- & 1.04 & 6x6x1 & 2 & 4 \\
        \scGWo/-BSE\cite{shi13} & 2.22 & 2.22 & 2.5 & 2.5 & --&0.63 & 15x15x1 & 6 & 8 \\
	\end{tabular*}
    \end{ruledtabular}
\end{table*}

Next, we consider the absorption spectrum of \MoS2. In order to calculate the excitation spectrum including electron-hole interactions, we solve the Bethe-Salpeter equation for the electron-hole amplitude\cite{rohlfing98,rohlfing00}. The matrix elements of the BSE Hamiltonian are first calculated on a 12x12x1 k-grid and subsequently interpolated onto a much finer 72x72x1 k-grid, which is then diagonalized to yield the exciton states and the resulting absorption spectrum.

\begin{figure}
    \includegraphics[width=246.0pt]{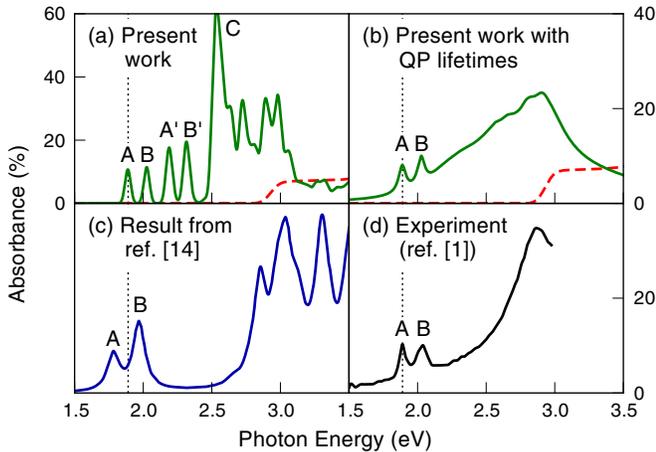}
    \caption{\label{fig:spectra} (Color online) (a) Absorption spectra of \MoS2 without (dashed red curve) and with (solid green curve) electron-hole interactions using a constant broadening of $20$~meV. (b) Same calculated data as in Fig.~\ref{fig:spectra}~(a), but using an \textit{ab initio} broadening based on the electron-phonon interactions\cite{marini08,li13}. (c) Previous \GoWo/ calculation (in arbitrary units). Note that the region between 2.2 and 2.8~eV is completely flat. (d) Experimental absorbance\cite{mak10}.}
\end{figure}

The absorption spectra for \MoS2 with and without electron-hole interactions are displayed in Fig.~\ref{fig:spectra}~(a). We find that there is a large exciton binding energy of 0.96~eV for the lowest energy exciton, and the first peak in the absorption spectrum is split into two peaks by spin-orbit coupling. We will follow the notation used in the experiment paper\cite{mak10} and refer to these two peaks as A and B and the next three lowest energy peaks we label as A', B', and C respectively. In our calculation, peaks A and B are located at 1.88 and 2.02~eV. This is in good agreement with experimental absorption peaks at 1.88 and 2.03~eV\cite{mak10}. A previous \GoWo/-BSE calculation\cite{ashwin12} also obtains agreement with experimental results for the positions of peaks A and B. However, we see multiple additional peaks in the absorption spectrum at photon energies between the energy of peak B at 2~eV and the quasiparticle gap at 2.8~eV, which were not seen before.

\begin{figure}
    \includegraphics[width=246.0pt]{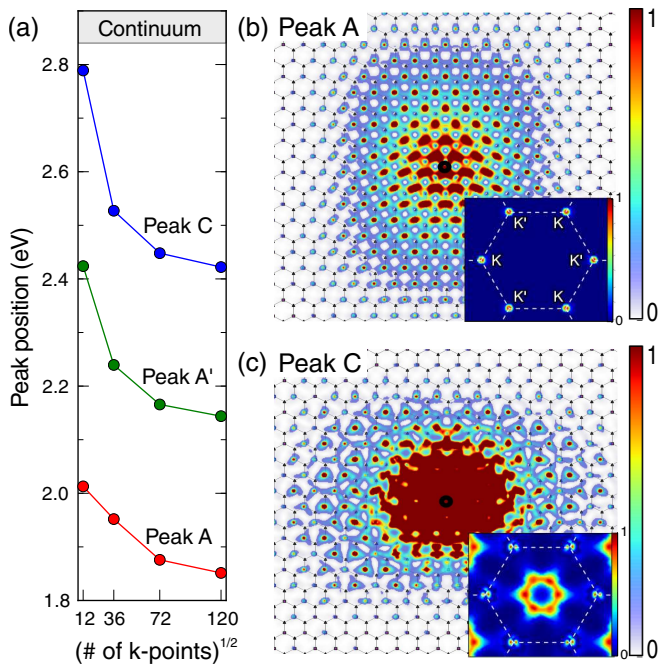}
    \caption{\label{fig:excitons} (Color online) (a) Convergence of the excitation energies of peaks A, A' and C as a function of the number of k-points. (b) Exciton corresponding to peak A in real space and in k-space (inset). The real-space plot is the modulus squared of the exciton wave function projected onto the plane with the hole (black circle) fixed near a Mo atom. Mo atoms are purple squares, and sulfur atoms are green triangles. (b) Real-space and k-space (inset) modulus squared of excitons forming peak C.}
\end{figure}

In order to obtain accurate absorption spectra, it is important to pay attention to the converge properties of the Bethe-Salpeter equation. There are two important factors that come into the BSE Hamiltonian: the number of Bloch bands and the number of k-points. We find that the spectrum converges quickly with the number of bands included, with just the 2 highest valence and 4 lowest conduction bands being sufficient. Going up to 7 valence and 8 conduction bands does not affect the optical spectrum up to $\omega\sim 3.5$~eV. However, we find that the accurate position of peaks A and B and the \textit{existence} of higher energy peaks are very sensitive to the number of k-points used in the BSE. As the number of k-points increases, there are numerous qualitative changes in the absorption spectrum (see Fig.~\ref{fig:excitons}~(a) and the Supplemental Material for details).

We sample up to 120x120 k-points and find that the peak positions and features are converged to better than $0.1$~eV only on a 72x72 k-grid, as shown in Fig.~\ref{fig:excitons}~(a). If we coarsen our k-grid from 72x72x1 to 12x12x1, the spectrum changes qualitatively, and we observe only one additional peak between 2.0 and 2.8~eV. Further decreasing k-point sampling to a 6x6x1 k-grid red-shifts the spectrum by as much as 1~eV. Thus, we believe that previous calculations\cite{ashwin12,shi13} are not converged with respect to k-point sampling, and the agreement between peak position for A and B and experimental results found in Ref.~\cite{ashwin12} is likely accidental. In Fig.~\ref{fig:spectra}, we emphasize that the additional bound exciton peaks that arise in the converged calculation are necessary to reproduce features in the experimental absorption spectrum above 2~eV.



Next, we show that the apparent discrepancy in the fine features between our calculated absorbance spectrum in Fig.~\ref{fig:spectra}~(a) and experiment\cite{mak10} in Fig.~\ref{fig:spectra}~(d) is actually a signature of lifetime effects due to electron-phonon interactions, which we can take into account. It was shown that the quasiparticle scattering rate in \MoS2 due to electron-phonon interactions is highly dependent on the quasiparticle energies, exhibiting a sudden increase when the quasi-electrons have enough energy to emit an optical phonon ($\sim 50$~meV)\cite{li13}. We incorporate this quasiparticle lifetime broadening in our absorption spectrum following Ref.~\cite{marini08}, and the result is plotted in Fig.~\ref{fig:spectra}~(b). We consider both emission and absorption of phonons at $T=300$~K, and we extrapolate the scattering rate for quasiparticle energies larger than those computed in Ref.~\cite{li13} (see Supplemental Material). With this approach, it is quite remarkable that the first two peaks (A, B) in Fig.~\ref{fig:spectra}~(b) remain relatively sharp, while the higher-energy peaks coalesce into a much broader blob, in good agreement with experiment, which shows large absorption in the energy range above 2~eV~\footnote{We found this procedure to be robust and it is not much dependent on our approximation to extrapolate the scattering rate, as discussed in the Supplemental Material.}. This feature cannot be explained by the existence of just the A and B states only. This also suggests that the optical properties of \MoS2 may be altered by tuning the electron-phonon interactions, such as by employing mechanical strain or dielectric screening. 

We look now at the character of the excitons giving rise to the low-energy peaks (A, B, A',B', and C) in the absorption spectrum. Peak A is made up of two energetically degenerate excitons formed, respectively, from free electron-hole transitons at the $K$ and $K'$ valleys (Fig.~\ref{fig:excitons}~(b) inset). The wave functions of these two excitons look identical in real space. In Fig.~\ref{fig:excitons}~(b), we show the modulus squared of the real-space exciton wave function when the hole is fixed near a Mo atom. The character of the exciton corresponding to peak A reflects the Mo d$_{z^{2}}$ orbitals of the states in the lowest \MoS2 conduction band. The envelope of the exciton wave function is nearly azimuthally symmetric, resembling the $1s$ state in a 2D hydrogenic model. The root mean square radius of the exciton in real space is 1~nm. Since spin is a good quantum number at $K$ and $K'$\cite{xiao12}, the spatial components of the spinor wave functions for the spin-orbit split bands are identical at $K$ and $K'$. Thus, since peaks A and B are made up of transitions between the spin-orbit split valence band and the lowest conduction band at $K$ and $K'$, the exciton wave functions corresponding to peak B is identical to that of peak A. The two next lowest energy peaks, near 2.2~eV, which we refer to as peaks A' and B', are also spin-orbit split states of excitons localized near the $K$ and $K'$ valleys in momentum space. These excitons have wave functions with a single node structure in k-space and are excited states of the exciton states forming peaks A and B (not shown). The excitons forming peaks A' and B' are six-fold degenerate, corresponding to the three degenerate 2s and 2p states in a 2D hydrogen model, multiplied by the two valleys of K and K'. There is also a cluster of six nearly-dark excitons near 2.35~eV (referred to as A", not shown), which are the second excited states of the excitons forming peak A, and a third set of excited states at 2.60~eV, which hybridizes slightly with the excitons forming peak C. We emphasize here that although we use the notations (e.g. ``2s'') of the 2D hydrogen model to refer to the states, our GW-BSE exciton spectrum is completely different from that of a standard 2D hydrogen model (see Fig. 3 in the Supplemental Material).

Next, we look at the character of the states forming the large peak C near 2.5~eV in Fig.~\ref{fig:spectra}~(a), which is plotted in Fig.~\ref{fig:excitons}~(c). We emphasize that this is the first time this peak has been seen in a theoretical or experimental study. The peak comes from six nearly-degenerate exciton states made from transitions between the highest valence band and the first three lowest conduction bands near, but not directly at $\Gamma$, reflecting the fact that the lowest transitions are degenerate at multiple points near $\Gamma$ (Fig.~\ref{fig:bandstructure}). When the sum of the k-space modulus squared of the six excitons is plotted (inset Fig.~\ref{fig:excitons}~(c), we see that it is indeed located in a ring with six-fold symmetry around the $\Gamma$ point, with some small contributions from points near $K$ and $K'$. This set of excitons has finely structured contributions from different parts of momentum space, which explains the need for a fine k-point sampling to converge the optical absorption spectrum, and may also explain why previous studies with coarse k-point sampling did not observe this peak. When these excitons are plotted in real space, with the hole fixed near a Mo atom, the electron has both Mo $d_{z^{2}}$ and S $p_{x}$ and $p_{y}$ character, with more S character near the hole. The real-space envelope of the C peak excitons has near azimuthal symmetry and a mean squared radius of 0.5~nm.

In conclusion, we have performed first-principles calculations of the quasiparticle band structure and excitonic properties of \MoS2 with many-body effects included. We identify and analyze the character of exciton peaks in the optical spectrum, many of which have not been experimentally or theoretically observed, and predict that the region in the optical spectrum between 2.2 and 2.8~eV is not featureless, but contains many bright and dark excitonic states which are broadened by electron-phonon interactions. This suggests that variations in temperature, dielectric screening, and mechanical strain could affect the optical absorption in this region of the spectrum. Also, the predicted features associated with the higher energy excitonic states may be enhanced and thus observed in experiment at reduced temperature and/or through modulation techniques such as electro-modulation spectroscopy. The large number and diverse character of excitons in the energy window between 2.2 and 2.8~eV suggests that \MoS2 is an ideal system for probing both inter-exciton transitions (e.g. between A and C) and intra-exciton transitions (e.g. between A, A', and A"), using methods such as pump-probe spectroscopy and two-photon experiments.

We thank M. Jain, A. Ramasubramaniam, and T. Cao for discussions. D. Y. Q. acknowledges support from the NSF Graduate Research Fellowship Grant No. DGE 1106400. F. H. J. acknowledges partial support from the Office of Naval Research under the MURI program. Structural study and the work on calculating the electron-phonon interaction effects on the optical spectra were supported by NSF Grant No. DMR10-1006184. The GW-BSE calculations were supported by the Theory Program at the Lawrence Berkeley National Lab through the Office of Basic Energy Sciences, U.S. Department of Energy under Contract No. DE-AC02-05CH11231. S. G. L. acknowledges support of a Simons Foundation Fellowship in Theoretical Physics. This research used resources of the National Energy Research Scientific Computing Center, which is supported by the Office of Science of the U.S. Department of Energy.
 

\bibliography{MoS2}
\bibliographystyle{apsrev4-1}

\end{document}